# Societal and Ethical Interactions with Nanotechnology ("SEIN") – An Introduction

DAVIS BAIRD* and TOM VOGT**

**ABSTRACT**

*Recent years have witnessed fundamental changes in the interrelationships between social and ethical norms and technological change. Public debate increasingly focuses on a more serious dialogue concerning the impact of technological change on social and ethical values. Nanotechnological development epitomizes this new dynamic in many ways, and its potential to catalyze enormous change make nanoscience and technology truly post-academic. In this article, Davis Baird and Tom Vogt identify six important issues tied to the continued development of nanotechnology: (1) environmental issues; (2) equitable issues relating to the possible emergence of a "nano-divide"; (3) legal, regulatory and insurance challenges; (4) privacy issues; (5) the interaction between nanomedicine and medical ethics, more traditionally; and (6) "hypertechnology," or the pace of nanotechnological change. Engaging all of nano's stakeholders—from various publics to entrepreneurs, industrialists, venture capitalists, scientists and engineers—demands establishing a context for dialog that is open, honest and attentive to interests and equity. So far, the nanotechnology community has been successful in responding to the concerns of others about the ethical and social challenges posed by nanotechnological development. Baird and Vogt conclude that continued efforts to build upon this emerging inclusive dialog are needed to ensure a mutually beneficial relationship between nanotechnology and society.*

## INTRODUCTION

In the past decade we have witnessed the beginning of a transformation in science, which is radically changing the research culture. This transformation is often referred to as a change from Mode-1 to Mode-2 knowledge production[1] or the emergence of post-academic science.[2] Nanoscience represents the first major scientific endeavor that embodies this transformation. The traditional scientific ethos had no place for ethics. As a matter of fact it was regarded as an external factor, which needed to be disregarded in the interest of 'neutrality'. The academic quest for new scientific knowledge takes precedence over all other considerations including ethical concerns. The ongoing cultural change occurring in science results in a stronger contextualization and embedding of science in society.[3] As a result, science is being questioned and challenged by the public sphere in an unprecedented manner resulting in disorientation, anger and condescending responses from many scientists. Post-academic science no longer denies, but faces the ethical dimensions and moral dilemmas and attempts to engage the public early on in a debate about the technological future. The National Nanotechnology Initiative ("NNI"), and in particular the way it has been pursued by the National Science Foundation ("NSF") and the Department of Energy ("DOE"), has required consideration of societal and ethical issues as part of all NSF/DOE centers funded to pursue nanotechnology, and has funded independent work on societal and ethical interactions with nanotechnology. In this sense, nano- science and technology is post-academic.

---

* USC Nanocenter, University of South Carolina
** Center for Functional Nanomaterials, Brookhaven National Laboratory, Upton, NY 11973-5000

[1] MICHAEL GIBBONS, ET AL., THE NEW PRODUCTION OF KNOWLEDGE (1994).
[2] JOHN ZIMAN, REAL SCIENCE (2000).

In the following, we briefly outline the most important societal and ethical issues tied to progress in nanoscience and technology. These include issues tied to (1) the environment; (2) equity, or a possible "nano-divide;" (3) legal, regulatory and insurance issues; (4) privacy; (5) issues associated with nanomedicine and medical ethics; (6) and issues tied more generally to the rapid advance of technology, what we call "hypertechnology." We close with some further remarks about how these issues are being approached—and need to be approached—in a multi-disciplinary way. This is further evidence that nanotechnology is a fully "mode-2" science.

I.  ENVIRONMENT

Environmental concerns are currently one of the biggest public concerns. Nanotechnolgy's promise is that it will provide new means for pollution remediation and less toxic ways to manufacture goods. However, latent toxicity is the flipside: nanosize materials are interesting because their physical and chemical properties differ so radically from bulk amounts of chemically the same compounds. Some of these differences are useful and wanted, but others have the potential to be less desirable. Moreover, these differences open the door to the need to re-examine and alter, if necessary, the regulatory framework for ensuring the safety of substances brought to the workplace or to the public. The materials safety data sheets ("MSDS") of carbon nanotubes and "buckyballs" are identical to the ones used for graphite despite the fact that they are three distinct elemental modifications of carbon with very different properties. This has been recognized as a regulatory deficiency, but a solution involves many stakeholders and will stretch existing legal frameworks and toxicology testing capabilities. Finally, the appearance of potential toxicities raises immediate problems for both assessing and communicating risk. Inadequate assessment and communication of potential hazards and risks could result in detrimental effects in the public arena and investment world for all of nanotechnology. If the public comes to associate the term, "nanotechnology," with toxic risk, association will tar other benign and beneficiary nanotechnologies. It is vital that we tackle the coupled scientific and legal problems quickly and effectively as they arise. Therefore, effective tools for a 'real-time technology assessment"[4] need to be developed.

II.  EQUITY

Equity concerns are equally fundamental, but even broader in scope. They encompass environmental concerns, for it is legitimate to worry about an inequitable distribution of risks and benefits in the development of and exposure to new materials. Workplace safety laws help ensure that workers do not shoulder an inequitable share of risk from exposure to toxic materials while owners and shareholders enjoy an inequitable share of benefits. Looking further into the future, various nanotechnologies are likely to create severe economic disruptions. We can foresee new technologies as a result of nanoscientific advances (i.e. solar energy, "hydrogen economy") that could impact the fortunes of those invested in and working for current technologies (i.e. oil). In a transformational period between a carbon and non-carbon based economy, inequity between urban and rural areas might emerge as an undesired consequence. A rural area might require initial subventions to avoid delays in the development of a hydrogen infrastructure, which could result if market forces were the only consideration. The shutdown of the petroleum infrastructure will lead to significant employment shifts in certain areas of the country, in particular, the Gulf coast. In general, oil-producing regions are more rural and alternative jobs are not as readily available. During the transformation to a 'carbon-free' hydrogen economy, new facilities need to be constructed leading to a 'boom' in some areas. Thus a migration of the 'old' and 'new' workforce will create regional winners and losers. Workforce displacement, industrial displacement and regional displacement producing a new "rust belt," all raise equity issues that need judicious examination. Questions concerning equity, while already daunting on a national scale, become even more complicated when considered in worldwide geo-political terms. One of the foremost questions asked is if nanotechnology will address the most urgent problems of developing countries (energy, clean water, food) or just accept a "global nanodivide" as we have come to live with a "digital and genetic divide".

---

[3] HELGA NOWOTNY, PETER SCOTT & MICHAEL GIBBONS, RE-THINKING SCIENCE (2001).
[4] David H. Guston & Daniel Sarewitz, *Real-Time Technology Assessment*, 24 TECH. IN SOC'Y 93 (2002), *available at* http://www.idehist.uu.se/personal/s_widmalm/STS%20Home/Resurser/Textmaterial/techassess.pdf.

### III. LEGAL, REGULATORY AND INSURANCE ISSUES

Already by now, it should be clear that various nanotechnologies are raising a large number of legal, regulatory and insurance related issues. Are we properly regulating new nano materials? How will our already strained health insurance system cope with human enhancements and precautionary medical interventions? The push to accelerate technology transfer is redefining previously stable institutions and institutional relationships, particularly with respect to the role of intellectual property in relationships between government, the university and industry. There are concerns about whether the current patent system is properly encouraging the development of new nanotechnologies. Nanotechnology is developing internationally, and there is a need to take steps to insure some commonly accepted ground rules, with respect to nomenclature, approaches to risk and patents and intellectual property rights.

We need to find novel and better ways to use the regulatory and insurance systems to distribute risk and benefit, while at the same time promoting research and entrepreneurial efforts. The electricity grid needs to become an "open source" into which consumers and a large number of independent energy producers can tap into and draw from. It is conceivable that an "energy web" will emerge where distribution and "supply and demand" self-organizes and frequent technology upgrades on the consumer end will allow *all* energy producing options (solar, wind, nuclear, biomass, geothermal, carbon-based) to "hook up"– such an energy revolution from 'general electric' to 'private electric' might be as far-reaching as what happened in computing with the advent of the PC and the world-wide web.

### IV. PRIVACY

Advances in information technology are already now augmenting concerns about privacy, but these will be severely heightened by likely developments in nanotechnology. Ubiquitous information technologies, connected to multiple and, in some cases invisible, sensors, communicating and exchanging information through increasingly common wireless networks raise possibilities for all kinds of privacy mischief. Health examples are striking. Embedded diagnostic devices in nanomedical applications that communicate results by wireless technology could make highly personal information widely available.

Public "medical outing" of personal health issues of politicians or celebrities and the criminal interference with such networks will present legal challenges. Changes in the kind of information that can be discovered with new diagnostics will have both insurance and personal consequences that will be difficult to resolve. But, health is not the only domain to consider. How do we redefine notions of privacy and public space when unobtrusive and even invisible sensors provide data about people in public spaces? What about private spaces?

## V.  MEDICAL ETHICS

Nanomedicine will raise—and already is raising—controversial issues tied to nanotechnological advances in life sciences. Just a few examples: Natural/artificial hybrids developed from tissue engineering, the use of engineered viruses, semi-autonomous diagnostic systems (e.g., lab-on-a-chip) that use wireless communication technology and expert systems to develop diagnoses. Natural/artificial hybrids and engineered viruses raise fear about introducing functionally novel materials into the human system. "Lab-on-a-chip" diagnostic systems raise apprehension about privacy and legal responsibility for mistaken diagnoses. Research topics beyond privacy (with their risks associated with new informational systems) include questions of discrimination; newly emerging concepts of "normality" and "pathology;" and the multiple issues raised by novel emergent diagnostic or therapeutic modalities to name just a few.

These developments, and many others, raise anxieties about how the doctor/patient relationship will develop in such a new world. We need to consider how the social configurations associated with nanomedicine will relate to current biomedical norms of practice, and how established medical ethical norms may be revised so that values integral to those norms can be sustained, while also allowing for the many advantages that a more 'personalized' nanomedicine may make possible.

Many developments in nanomedicine are likely to hit in the near to medium term. Further out, is work on more radical interventions for human enhancement, including brain/device computation and communication interfaces. Work by the "converging technologies" program, or "NBIC" suggest fairly radical changes ahead:

"For example, it is not possible to see a century into the future, but it may be that humanity would become like a single, distributed and interconnected "brain" based in new core pathways of society. This will be an enhancement to the productivity and independence of individuals, giving them greater opportunities to achieve personal goals."[5]

Such change—if and when it may come about—would severely tax our culture's institutions and traditions. We need to examine very carefully what kind of society we might be building here.

## VI.  HYPERTECHNOLOGY

Ray Kurzweil has focused much of his discourse on the accelerating pace at which technological change comes upon us.[6] Beyond "Moore's Law," he invokes double-exponential rates of change for a large number of metrics.[7] Built into the NNI, is language pushing for a more rapid development of nanotechnological progress. Such claims and aims for what we call "hypertechnology" demand consideration both about the facts of the matter *and about their perception*. Part of the point of engaging in research on societal and ethical interactions with nanotechnology is to help manage and promote a smooth technological transition to a nano-embedded world. This will require careful thought about the possibility of rapid technological change and—again—the possible perception of rapid technological change. Society will adapt to new technologies by adapting these technologies to itself. How, then, will society—or, better put, multiple and diverse societies—understand and adapt nanoenhanced selves and societies to themselves?

---

[5] NAT'L SCI. FOUND. (NSF), *Converging Technologies for Improving Human Performance: Nanotechnology, Biotechnology, Information Technology and Cognitive Science*, (Mihail C. Roco & William S. Bainbridge, eds.) (2002), *available at* http://www.wtec.org/ConvergingTechnologies/Report/NBIC_report.pdf
[6] RAY KURZWEIL, THE AGE OF SPIRITUAL MACHINES (1999)
[7] *Id*.

## VII. TRANSDISCIPLINARITY

As mentioned in the beginning, a new research paradigm is emerging: teams of interdisciplinary—even transdisciplinary—experts tackling problems with ethical consequences for society at large are replacing "lonely seekers of truth," who define their own problems and ignore societal and ethical dimensions. Consider work on aging, for example. Some researchers at the border of nanomedicine and human biology have suggested that we are on the verge of significantly gaining control over the aging process and significantly extending the average lifespan.[8] Of course, one important question to ask concerns the evidence for such claims. If true, however, the implications for society are staggering. One can immediately see—and seek more careful quantitative information concerning—implications for population size, the social security system, and numerous other metrics. But, arguably, the qualitative impact on the social fabric would be much more profound. Were the average lifespan 110 or 120 (to pick a number), how would family structures respond with five, six or seven living generations? How would the institution of marriage respond to 100-year "double platinum" anniversaries? How would we think about and value 19-year-old lives lost to car accidents or a preemptive war when they might have gone on to live 120 years? Does this change how people think about "risky behavior?" Would such an extension of lifespan affect how we think about the importance of being mortal? The point here is that we cannot be satisfied with work that does not integrate quantitative research with more qualitative or "human-meaning-related" (or "hermeneutic") research. Societal interactions do not reside in nice tidy (university-driven) disciplinary boxes, rather they grow "between the cracks."

## CONCLUSION

One final point about what we take to be the most essential need in moving ahead with addressing societal and ethical interactions with nanotechnology. In the past, most technology assessment, and most efforts in evaluating ethical, legal and social implications of new technologies, "black-boxed" the science and engineering involved, and focused exclusively on "impacts." Technical details can be difficult to understand and communicate to other stakeholders, and a "quasi-scientific" notion of expertise can seem to support the idea that the technical details should be left to the scientists and engineers, while social and ethical details should be the province of humanist, social science, legal and policy "experts." This is all wrong-headed. Societal impacts cannot be controlled and tweaked the way a microscopist might improve an electron microscope. Scientists, engineers, and all of the many stakeholders in our joint socio-nano-technological future, need to engage in multi-directional discussions about societal values, needs, scientific/engineering prospects and probabilities. We envision such multi-directional discussions along the lines of a huge version of an old-fashioned town meeting, where we are all collectively constructing the "nano society" of the future. There are models for the public participation in science and technology

---

[8] Michael R. Rose, *The Science of Staying Young*, 14 SCI. AM. 23 (2004) (special edition), *available for purchase at* http://www.sciam.com/special/index.cfm?issueid=23&sc=I100381.

AUTHOR NAME GOES HERE

decision making such as community-based research, the Danish Consensus Conferences, Citizen Panels, Scenario Workshops;[9] however, they need to be genuinely embedded within the fabric of our legal and political system.

We need to provide spaces for such inclusive stakeholder dialogues. Three important characteristics to success here are awareness, understanding and trust. These are not unidirectional concepts. Currently, there is limited public awareness of nanotechnology and little understanding of nanotechnology. But, vice versa, there is limited awareness and understanding on the part of scientists and engineers about how different publics think about and respond to new technologies. The unanticipated public rejection of "Frankenfoods" supports this point. It is easy to think that a lack of knowledge is the problem—and we don't mean to suggest that it isn't a problem—but very likely it is trust and sensitivity to power, conflicts of interest and equity that lie at the heart of the problem. Engaging all of nano's stakeholders—from various publics to entrepreneurs, industrialists, venture capitalists, scientists and engineers—demands establishing a context for dialog that is open, honest and attentive to interests and equity. Our impression is that the "nano community" has done pretty well so far. We have heard that Pat Mooney of the ETC Group—an advocacy group that early on raised concerns about nanotechnology—is impressed so far with how responsive the "nano community" has been to concerns. We can and should take advantage of this early success and we can and should take advantage of how early in the game we are. Success in building inclusive dialog is possible.

---

[9] Jill Chopyak & Peter Levesque, *Public Participation in Science and Technology Decision Making*, 24 TECH. IN SOC'Y 155 (2002).